\documentclass[a4paper,11pt]{article}
\pdfoutput=1

\usepackage{jheppub} 

\usepackage{graphicx}
\usepackage{amsmath}
\usepackage{amssymb}
\usepackage{hyperref}
\usepackage{color}
\usepackage{transparent}
\usepackage{booktabs}

\makeatletter
\let\old@fpheader\@fpheader
\renewcommand{\@fpheader}{\old@fpheader\hfill
MPP-2016-326}
\makeatother

\title{Supersymmetric Casimir Energy and $\mathrm{SL(3,\mathbb{Z})}$ Transformations}

\author[a,b]{Frederic Br\"unner,}
\author[b,c]{Diego Regalado,}
\author[d,e]{and Vyacheslav P. Spiridonov}

\affiliation[a]{Institut f\"ur Theoretische Physik, Technische Universit\"at Wien\\
        Wiedner Hauptstra\ss e 8-10, A-1040 Vienna, Austria}

\affiliation[b]{Max-Planck-Institut f\"ur Physik (Werner-Heisenberg-Institut)\\
      F\"ohringer Ring 6, D-80805 Munich, Germany}

\affiliation[c]{Institute for Theoretical Physics and Center for Extreme Matter and Emergent Phenomena\\
     Utrecht University, Leuvenlaan 4, 3584 CE Utrecht, The Netherlands}

\affiliation[d]{Laboratory of Theoretical Physics, JINR\\ J. Curie str. 6,
Dubna, Moscow region, 141980, Russia}

\affiliation[e]{Laboratory of Mirror Symmetry, NRU HSE\\ 6 Usacheva str.,
Moscow, 119048, Russia}

\emailAdd{bruenner@hep.itp.tuwien.ac.at}
\emailAdd{regalado@mpp.mpg.de}
\emailAdd{spiridon@theor.jinr.ru}

\abstract{We provide a recipe to extract the supersymmetric Casimir energy of theories defined on primary Hopf surfaces directly from the superconformal index. It involves an $\mathrm{SL(3,\mathbb{Z})}$ transformation acting on the complex structure moduli of the background geometry. In particular, the known relation between Casimir energy, index and partition function emerges naturally from this framework, allowing rewriting of the latter as a modified elliptic hypergeometric integral. We show this explicitly for $\mathcal{N}=1$ SQCD and $\mathcal{N}=4$ supersymmetric Yang-Mills theory for all classical gauge groups, and conjecture that it holds more generally. We also use our method to derive an expression for the Casimir energy of the nonlagrangian $\mathcal{N}=2$ SCFT with $\mathrm{E_6}$ flavour symmetry. Furthermore, we predict an expression for Casimir energy of the $\mathcal{N}=1$ $\mathrm{SP(2N)}$ theory with $\mathrm{SU(8)\times U(1)}$ flavour symmetry that is part of a multiple duality network, and for the doubled $\mathcal{N}=1$ theory with enhanced $\mathrm{E}_7$ flavour symmetry. }

\begin{document}
\maketitle

\section{Introduction}

Supersymmetric gauge theories on curved spacetime manifolds have been studied intensely in recent years. The rise of localization as a tool to compute exact partition functions, especially in four-dimensional
theories \cite{Nekrasov}, has led to many interesting results and allowed the field to flourish (see \cite{pestun} for a recent collection of reviews). Another quantity of interest is the superconformal index \cite{KMMR,R}, which can be defined for theories on Hopf surfaces, complex manifolds with $\mathrm{S^1\times S^3}$ topology. It is not affected by supersymmetry-preserving deformations and as such is invariant under the renormalization group flow. It receives contributions only from short representations and has served as a powerful check for many dualities. By applying the technique of localization \cite{ACM,ACDKLM} (see also \cite{CS,ALS}), it was found that the index $\mathcal{I}_{\mathrm{SC}}$ is related to the partition function $\mathcal{Z}_{\mathrm{SUSY}}$ by

\begin{equation}\label{rel}
\mathcal{Z}_{\mathrm{SUSY}}=e^{-\beta E_{\mathrm{Casimir}}}\;\mathcal{I}_{\mathrm{SC}},
\end{equation}

\noindent where $\beta$ determines the size of the $\mathrm{S^1}$ submanifold and $E_{\mathrm{Casimir}}$ is the supersymmetric Casimir energy. The latter determines the leading order behavior of the partition function in the $\beta\rightarrow\infty$ limit and is given in even dimensions by an equivariant integral
of the anomaly polynomial \cite{BBK}. It has also been investigated in
the context of holography in \cite{GCMS}.

Earlier, in an independent long term study inspired by quantum mechanical considerations,
the third named author discovered the elliptic hypergeometric integrals \cite{spi},
the top level special functions of hypergeometric type. They generalize the plain hypergeometric
functions and their $q$-analogs by adding one more (elliptic) deformation parameter $p$,
also called a basic parameter. The original physical
motivation was justified by application of these integrals in relativistic integrable many-body
models \cite{spi2007}. The relation to supersymmetric field theories was discovered in \cite{DO}.
More precisely, superconformal indices appeared to be identical to certain elliptic hypergeometric
integrals whose symmetry transformations rigorously confirm Seiberg dualities \cite{Seiberg}
in the sector of BPS states. From \eqref{rel} it follows that the same integrals
describe supersymmetric partition functions of four-dimensional field theories.
As observed in \cite{conm}, the superconformal indices of quiver theories represent partition
functions of solvable two-dimensional lattice models. An explanation of this fact was given in
\cite{Yagi} through the relation to two-dimensional topological field theories of \cite{GPRR}.
A recent survey of this subject with references to relevant papers is given in \cite{RR2016}.
All of this shows the universal relevance of elliptic hypergeometric functions for
applications both in physics and mathematics.

One of the key checks of the validity of Seiberg dualities was a verification of 't Hooft anomaly matching
conditions for dual theories \cite{Seiberg}. As shown in \cite{SV}, $\mathrm{SL(3,\mathbb{Z})}$ transformations are related to these matching conditions, since they reproduce all
anomaly coefficients through the cocycle phase factor emerging in the corresponding transformation
rule for the elliptic gamma function \cite{FV}.

The partition function of a two-dimensional conformal field theory on a torus is invariant under
the modular group $\mathrm{SL(2,\mathbb{Z})}$. The Casimir energy determines its leading order behavior
as well, in complete analogy to the four-dimensional case.
However, it is easy to see that the partition function of Eq.~(\ref{rel}) is not modular invariant. The same holds true for the superconformal index. E.g., in the
so-called Schur limit degeneration of $\mathcal{N}=2$ index it turns out \cite{Razamat} that
$\mathrm{SL(2,\mathbb{Z})}$ modular invariance
is obtained only after modifying the index. Geometrically, the modular transformation acts on one fugacity
parameter of the reduced index and corresponds to the usual action on a torus.
The general superconformal index depends on a larger number of parameters. As such, it admits the action of the group $\mathrm{SL(3,\mathbb{Z})}$ transforming the complex structure moduli of the Hopf surface
in a particular way, affecting $\beta$ and the other ``isometry parameters''.

In this article, we study the transformation behavior of the superconformal index under $\mathrm{SL(3,\mathbb{Z})}$ in more detail. Our first observation is that different $\mathrm{SL(3,\mathbb{Z})}$ transformations act on complex structure moduli differently. While none leave them invariant, some map real ones into real, some real into complex parameters. In particular, there exists a transformation that partially relates the high and low ``temperature''
\footnote{This is not really a temperature, since the boundary conditions for fermions on the $S^1$ are periodic.} asymptotics of the indices.
The high temperature limit, when both basic parameters $p$ and $q$ tend to 1, maps indices to the hyperbolic
integrals describing three-dimensional partition functions \cite{DSV}, while the low temperature limit,
when $p$ and $q$ go to zero, degenerates the indices to Hilbert series
counting the gauge invariant operators \cite{SV2}. The intermediate case, when only one of the
 base parameters $p$ or $q$ goes to zero, reduces to the Macdonald theory -- a relation which was noticed
first in \cite{SV3} and later developed in detail in \cite{GRRY1}.

Furthermore, we observe that the structure of the right hand side of Eq.~(\ref{rel}) arises as a consequence of the same $\mathrm{SL(3,\mathbb{Z})}$ transformation. This is because the elliptic gamma functions in the transformed superconformal index can be rewritten as a product of the modified elliptic gamma functions and exponentials consisting of the Bernoulli polynomials. The leading order contribution from those polynomials in the $\tilde{\beta}\rightarrow\infty$ limit, $\tilde{\beta}$ being the circle size of the transformed background, agrees precisely with the corresponding supersymmetric Casimir energy. As a direct consequence of Eq.~$(\ref{rel})$, the partition function can be written in terms of a modified elliptic hypergeometric integral. We demonstrate this explicitly for $\mathcal{N}=1$ SQCD and $\mathcal{N}=4$ supersymmetric Yang-Mills theory for all classical gauge groups. We also apply this technique to the superconformal index of $\mathcal{N}=2$ nonlagrangian theory with $\mathrm{E_6}$ flavour symmetry that arises in the context of Argyres-Seiberg duality \cite{MN,AS,GRRY2}.

We use the $\mathrm{SL(3,\mathbb{Z})}$ transformation to predict the supersymmetric Casimir energy of the $\mathcal{N}=1$ $\mathrm{SP(2N)}$ theory with $\mathrm{SU(8)\times U(1)}$ flavour symmetry that was studied in \cite{SV4} and that is part of a larger duality network. The other prediction we make concerns the Casimir energy of the $\mathcal{N}=1$ theory with enhanced $\mathrm{E_7}$ flavour symmetry constructed in \cite{DG}, and we also discuss a related 6d/4d theory. 

The article is organized as follows: in Section 2, we define the $\mathrm{SL(3,\mathbb{Z})}$ transformations and explain how they act on the kernels of elliptic hypergeometric integrals, the basic building blocks of superconformal indices. We also recall important facts about the Hopf surfaces. In Section 3, we introduce the superconformal index, study its behavior under $\mathrm{SL(3,\mathbb{Z})}$ transformations and show how the Casimir energy can be extracted, leading to Eq.~(\ref{rel}). In Section 4, we confirm this scheme explicitly for $\mathcal{N}=1$ SQCD, $\mathcal{N}=4$ SYM, $\mathcal{N}=2$ $\mathrm{E_6}$, and give our predictions regarding the $\mathcal{N}=1$ $\mathrm{SP(2N)}$ theory with $\mathrm{SU(8)\times U(1)}$ flavour symmetry, the theory with enhanced $\mathrm{E_7}$ flavour symmetry and the 6d/4d model.  We conclude and give a list of open questions in Section 5.

\section{Mathematical preliminaries}

In this section, we summarize mathematical statements required in the later parts of the paper.

\subsection{Elliptic hypergeometric functions}

Superconformal indices can be written as contour integrals of a product of elliptic gamma functions \cite{DO},
forming the elliptic hypergeometric integrals  \cite{spi},
a fact that has lead to a very fruitful interrelation between mathematics and physics. Mathematical properties of elliptic hypergeometric
integrals confirm Seiberg dualities by showing that they have identical sets of BPS states \cite{DO,SV2}.
Vice versa, known physical dualities lead to large number of new conjectural mathematical identities requiring rigorous proofs \cite{SV2}. In the following, we will introduce
these special functions and discuss their behavior under $\mathrm{SL(3,\mathbb{Z})}$ transformations.

The elliptic gamma function is the unique (up to the multiplication by a constant) meromorphic solution
to the finite difference equations
\begin{align}\label{finite}
f(u+\omega_1)&=\theta(z;p) f(u),\nonumber\\
f(u+\omega_3)&=\theta(z;q) f(u),\nonumber\\
f(u+\omega_2)&=f(u)
\end{align}

\noindent with $z=\mathrm{exp}(2\pi i u/\omega_2)$, $z\in \mathbb{C}^*$, and incommensurate
$\omega_j\in\mathbb{C}$. Here
$$
\theta(z;p)=(z;p)_\infty(pz^{-1};p)_\infty,\qquad (z;p)_\infty=\prod_{j=0}^\infty(1-zp^j),
$$
 is the Jacobi theta function. The bases $p$ and $q$ with $|p|,|q|<1$ are related to the complex parameters $\omega_i$ by $p=\mathrm{exp}(2\pi i \omega_3/\omega_2)$  and $q=\mathrm{exp}(2\pi i\omega_1/\omega_2)$. An explicit form of the elliptic gamma function $f(u)=\Gamma(z;p,q)$ is given by the infinite product

\begin{equation}
\Gamma(z;p,q)=\prod_{i,j=0}^\infty \frac{1-z^{-1}p^{i+1}q^{j+1}}{1-zp^i q^j}.
\end{equation}

The elliptic hypergeometric integrals defining the transcendental elliptic hypergeometric
functions are formed as contour integrals of particular products of $\Gamma(z;p,q)$
with special choice of the arguments $z$. The key characteristic property of these integrals
 \cite{spitheta} is that their integrand functions are defined as solutions of the first order finite-difference
 equations in the integration variables with the coefficients given by elliptic functions
 (i.e., the meromorphic double periodic functions).

 As shown in \cite{spitheta} there is another solution to the first line equation in Eq.~(\ref{finite}),
 such that the other equations are modified to

\begin{align}
f(u+\omega_2)&=\theta(e^{2\pi i u/\omega_1};r) f(u),\nonumber\\
f(u+\omega_3)&=e^{-\pi i B_{2,2}(u,\omega_1,\omega_2)}f(u),
\end{align}

\noindent where $r=\mathrm{exp}(2\pi i \omega_3/\omega_1)$ is an additional base, and $B_{2,2}$ is a second order Bernoulli polynomial given by the expression

\begin{equation}
B_{2,2}(u,\omega_1,\omega_2)=\frac{u^2}{\omega_1\omega_2}-\frac{u}{\omega_1}-\frac{u}{\omega_2}+\frac{\omega_1}{6\omega_2}+\frac{\omega_2}{6\omega_1}+\frac{1}{2}.
\end{equation}

\noindent This solution was called the \emph{modified elliptic gamma function}
$f(u)=\mathcal{G}(u;\omega)$, with $\omega=(\omega_1,\omega_2,\omega_3)$.
For $|p|, |q|<1$ it is related to the standard elliptic gamma function by

\begin{equation}\label{GMOD1}
\mathcal{G}(u;\omega)=\Gamma(r e^{-2\pi i u/\omega_1};\tilde{q},r)
\Gamma(e^{2\pi i u/\omega_2};p,q),
\end{equation}

\noindent with $\tilde{q}=\mathrm{exp}(-2\pi i\omega_2/\omega_1)$.
As follows from an identity derived in \cite{FV}, the modified elliptic
gamma function can be rewritten as

\begin{equation}\label{gmod}
\mathcal{G}(u;\omega)=e^{-\frac{\pi i}{3}B_{3,3}(u,w)}\Gamma(e^{-2\pi i u/\omega_3};\tilde{r},\tilde{p}),
\end{equation}

\noindent where $\tilde{r}=\mathrm{exp}(-2\pi i\omega_1/\omega_3)$, $\tilde{p}=\mathrm{exp}(-2\pi i\omega_2/\omega_3)$ and $B_{3,3}(u,\omega)$ is a third order Bernoulli polynomial given by

\begin{equation}
B_{3,3}(u,\omega)=
\frac{1}{\omega_1\omega_2\omega_3}(u-\frac{1}{2}\sum_{k=1}^3\omega_k)((u-\frac{1}{2}\sum_{k=1}^3\omega_k)^2-\frac{1}{4}\sum_{k}^3\omega_k^2).
\end{equation}

\noindent It is remarkable that $\mathcal{G}(u;\omega)$ remains a well defined meromorphic
function of $u$ even when the base $q$ lies on the unit circle, $|q|=1$, which is easily seen
from the representation \eqref{gmod}.

A key reason for introduction of the function $\mathcal{G}(u;\omega)$ as an
additional gamma function is the fact that many useful identities for
elliptic hypergeometric integrals are derived using only the first equation
in the set \eqref{finite}. Therefore it is natural to expect that there should exist
analogous identities formulated in terms of the function $\mathcal{G}(u;\omega)$,
which, in contrast to the original relations, will be well defined for $|q|=1$.

In section 4.3 we will need a double elliptic gamma function, given by

\begin{equation}
\Gamma(z;p,q,t)=\prod_{i,j,k=0}^\infty (1-zp^i q^j t^k)(1-z^{-1}p^{i+1}q^{j+1}t^{k+1}),
\end{equation}

\noindent where $t=\mathrm{exp}(2\pi i \omega_4/\omega_2)$ with $|t|<1$.
It also possesses a modified version in a similar spirit as that of the
original elliptic gamma function. This modified double elliptic gamma function
is defined by \cite{LMP2014}

\begin{equation}\label{doublegmod}
\mathcal{G}(u;\omega_1,\ldots,\omega_4)=\frac{\Gamma(e^{2\pi i u/\omega_2};q,p,t)}
{\Gamma(\tilde{q}e^{2\pi i u/\omega_1};\tilde{q},r,s)},
\end{equation}

\noindent with $s=\mathrm{exp}(2\pi i \omega_4/\omega_1)$. The vector $\omega$
now also includes $\omega_4$. There exists an analog of Eq. (\ref{gmod}), namely the relation

\begin{equation}\label{gmod2}
\mathcal{G}(u;\omega_1,\ldots,\omega_4)=e^{-\frac{\pi i}{12}B_{4,4}}
\frac{\Gamma(e^{-2\pi i u/\omega_3};\tilde{p},\tilde{r},\tilde{w})}
{\Gamma(we^{-2\pi i u/\omega_4};\tilde{s},\tilde{t},w)},
\end{equation}

\noindent with $w=\mathrm{exp}(2\pi i \omega_3/\omega_4)$ and
$\tilde{w}=\mathrm{exp}(-2\pi i \omega_4/\omega_3)$. The Bernoulli
polynomial $B_{4,4}\equiv B_{4,4}(u,\omega)$ is given by

\begin{equation}
B_{4,4}=\frac{1}{\omega_1\omega_2\omega_3\omega_4}\left[\left((u-\frac{1}{2}\sum_{k=1}^4\omega_k)^2-\frac{1}{4}\sum_{k=1}^4\omega_k^2\right)^2-\frac{1}{30}\sum_{k=1}^4\omega_k^4-\frac{1}{12}\sum_{1\leq j<k\leq 4}\omega_j^2\omega_k^2\right].
\end{equation}
As it is again easy to see, the function $\mathcal{G}(u;\omega_1,\ldots,\omega_4)$ is well
defined when the base $q$ lies on the unit circle, $|q|=1$.

For the description of the superconformal index, we also need the Euler function $\phi(q)$, which can be expressed in terms of the $q$-Pochhammer symbol as
$$
\phi(q)=(q;q)_\infty.
$$

 In \cite{SV}, it was discovered that `t Hooft anomaly matching conditions for Seiberg duality in $\mathcal{N}=1$
 theories are related to $\mathrm{SL(3,\mathbb{Z})}$ transformations acting on the homogeneous coordinates $\omega$ by

\begin{equation}\label{transformation}
\omega=(\omega_1,\omega_2,\omega_3)\longrightarrow \tilde{\omega}=(\omega_1,-\omega_3,\omega_2).
\end{equation}

\noindent Under this transformation, the bases $p,q,r$ change as
$p\rightarrow\tilde{p}$, $q\rightarrow\tilde{r}$ and $r\rightarrow\tilde{q}$.
For the elliptic gamma function, we get

\begin{align}
\Gamma(e^{2\pi i u/\omega_2};p,q)&\longrightarrow \Gamma(e^{-2\pi i u/\omega_3};\tilde{p},\tilde{r})\nonumber\\
&=e^{\frac{\pi i}{3}B_{3,3}(u,w)}\mathcal{G}(u;\omega),
\end{align}

\noindent as follows from Eq.~(\ref{gmod}). The transformation law of the Euler function follows from
the transformation properties of the Dedekind eta-function, and it is given by

\begin{equation}
\phi(e^{-2\pi i \frac{\omega_2}{\omega_1}})=\left(-i\frac{\omega_1}{\omega_2}\right)^{1/2}e^{\frac{\pi i}{12}(\frac{\omega_1}{\omega_2}+\frac{\omega_2}{\omega_1})}\phi(e^{2\pi i \frac{\omega_1}{\omega_2}}),
\end{equation}
where we assume that $\sqrt{-i}=e^{-\frac{\pi i}{4}}$. For the modified double
elliptic gamma function, the modular transformations are more involved. The bases $q, p, t$ in of the double elliptic gamma function in the numerator in \eqref{doublegmod}
change to $\tilde{p}, \tilde{r}, \tilde{w}$ in \eqref{gmod2}
corresponding to the transformation
$(\omega_2,\omega_3)\to (-\omega_3,\omega_2)$. However, for the function in the denominator one has the changes $\tilde q, r, s \to \tilde{t}, w, \tilde{s}$
corresponding to another $\mathrm{SL(2,\mathbb{Z})}$ subgroup action
$(\omega_4,\omega_1)\to (-\omega_1,\omega_4)$.

\subsection{Hopf surfaces}

In this section, we discuss basic facts about Hopf surfaces as described in \cite{ACM} and \cite{MS} (see also \cite{CDFK}). This is needed to understand the action of $\mathrm{SL(3,\mathbb{Z})}$ on the superconformal index geometrically. To avoid confusion, we mostly stick to the notation of \cite{ACM}.

One of the possible background geometries that preserve supersymmetry is $\mathrm{S^1\times S^3}$ \cite{FS}. In this case it is possible to have two complex Killing spinors of opposite R-charge, a requirement for $\mathcal{N}=1$ supersymmetry. We will restrict ourselves to \emph{primary Hopf surfaces}, which possess a fundamental group isomorphic to $\mathbb{Z}$. All primary Hopf surfaces are diffeomorphic to $\mathrm{S^1\times S^3}$. More general Hopf surfaces arise as quotients by specific finite groups. A prerequisite for the existence of two complex Killing spinors of opposite R-charge is that the manifold admits two complex structures of opposite orientation. Such primary Hopf surfaces are given by quotients of $\mathbb{C}^2-(0,0)$ of the form

\begin{equation}\label{hopf}
(z_1^{\pm},z_2^{\pm})\sim (p_\pm z_1^{\pm},q_{\pm} z_2^{\pm}),
\end{equation}

\noindent where $p_{\pm},q_{\pm}$ are complex structure moduli with $0<|p_{\pm}|\leq|q_{\pm}|<1$, and $z_1^\pm,z_2^\pm$ are complex coordinates. The signs $+$ and $-$ refer to the complex structures $I_+$ and $I_-$, respectively, and since we require them to be of opposite orientation with respect to each other, the manifold is said to possess ambihermitian structure.

To study Killing spinors on primary Hopf surfaces explicitly, a non-singular metric compatible with integrable complex structures was introduced in \cite{ACM}. This metric admits a complex Killing vector $K$ that satisfies $K_\mu K^\mu=0$ and commutes with its own complex conjugate. As a consequence, there exist two Killing spinors. Furthermore, the existence of an additional real Killing vector is assumed, leading to a $\mathrm{U(1)^3}$ isometry group. The metric is given by

\begin{equation}
ds^2=\Omega(\rho)^2 d\tau^2+f^2(\rho)d\rho^2+m_{IJ}(\rho)d\varphi_I d\varphi_J,
\end{equation}

\noindent with $I,J=1,2$, where $\tau\sim\tau+\beta$ parametrizes the $S^1$ part, while $\rho,\varphi_1,\varphi_2$ for $0\leq \rho\leq 1$, $0\leq \varphi_1\leq 2\pi$ and $0\leq \varphi_2\leq  2\pi$ are coordinates on the $S^3$. We describe the $S^3$ in terms of a torus fibration over the interval $0\leq\rho\leq1$, where $m_{IJ}(\rho)$ are the positive definite metric components of the torus, and $f(\rho),\Omega(\rho)>0$. The complex Killing vector is given by

\begin{equation}
K=\frac{1}{2}\left(b_1\frac{\partial}{\partial\varphi_1}+b_2\frac{\partial}{\partial\varphi_2}-i\frac{\partial}{\partial\tau}\right),
\end{equation}

\noindent with $b_1$ and $b_2$ being two real parameters. They are related to the complex structure moduli by

\begin{equation}
p_\pm=e^{\pm\beta|b_1|},\;\;\;\;\;\;\;\;\;\;\;\;\;\;\;q_\pm=e^{\pm\beta|b_2|}.
\end{equation}

\noindent While these parameters are real for a direct product metric like the one given above, for a non-direct product metric, they will in general be complex. The use of the letters $p$ and $q$ both in section 2 and section 3 is not a mere coincidence.

\section{The superconformal index and $\mathrm{SL(3,\mathbb{Z})}$}

In this section, we discuss the $\mathrm{SL(3,\mathbb{Z})}$ transformation properties of the superconformal index and present the main ideas of the article.

\subsection{The index}

The general expression for the superconformal index of an $\mathcal{N}=1$ theory defined on a primary Hopf surface is given by \cite{KMMR,R}

\begin{equation}\label{index0}
\mathcal{I}=\mathrm{Tr}(-1)^{\mathcal{F}}e^{-\gamma H}p^{\frac{R}{2}+J_R+J_L}q^{\frac{R}{2}+J_R-J_L}
\prod_{i} y_i^{F_i}\!\!,
\end{equation}

\noindent where $\mathcal{F}$ is the fermion number, $R$ is the $R$-charge, $J_L$ and $J_R$ are the Cartan generators of the rotation group $\mathrm{SU(2)_L\times SU(2)_R}$, and $F_j$ are maximal torus generators of the flavor group. The index only receives contributions from states with $H=E-2J_L-\frac{3}{2}R=0$, $E$ being the energy, and is independent of the chemical potential $\gamma$. While the parameters $y_i$ are fugacities for the flavor group, the basic parameters $p$ and $q$ are complex structure moduli of the Hopf surface and are given by

\begin{equation}\label{moduli}
 p=p_{-},\;\;\;\;\;\;\;\;\;\;\;\;\;\;\; q=q_{-}.
\end{equation}

For Lagrangian theories, its generic form is given by an integral over the gauge group:

\begin{equation}\label{index}
\mathcal{I}(p,q,y)=\int_G d\mu(z)\;\mathrm{exp}\left(\sum_{n=1}^\infty\frac{1}{n}i(p^n,q^n,y^n,z^n)\right),
\end{equation}

\noindent where $y$ and $z$ schematically denote all possible flavor or gauge
fugacities. The single particle index $i(p,q,y,z)$ only depends on the characters
of the representations of fields. It was shown that for many theories, this integral
can be rewritten in terms of elliptic hypergeometric functions that depend on
the complex structure moduli $p$ and $q$ as described in section 2. The fugacities will in general satisfy 
a constraint relation that is required for the consistency of the integral, and is related to the absence 
of gauge anomalies. 
For Seiberg dual theories the indices \eqref{index} coincide despite of
having quite different formal expressions.
This was shown first for $\mathcal{N}=1$ SQCD in \cite{DO}. For brevity,
we will not show how this comes about explicitly, and refer to \cite{SV2},
where many examples can be found.

\subsection{Mapping Hopf surfaces and asymptotics}

As described in the previous sections, the superconformal index for a theory
 defined on a primary Hopf surface depends on its complex structure moduli,
which arise as arguments of elliptic gamma functions. In order to see how
the $\mathrm{SL(3,\mathbb{Z})}$ transformations described in section 2.1 act
on them, we need to make the identifications $2\pi i \omega_3/\omega_2=-\beta|b_1|$
and $2\pi i \omega_1/\omega_2=-\beta|b_2|$. This is achieved by setting
$\omega_1=i|b_2|$, $\omega_3=i|b_1|$ and $\beta=2\pi/\omega_2$. Note that one
has $\omega_3/\omega_1\geq 0$, i.e. $|r|=1$.
Applying the transformation of Eq.~(\ref{transformation}) to these quantities, we get

\begin{equation}
p=e^{-\beta|b_1|}\longrightarrow \tilde{p}=e^{-2\pi i \frac{\omega_2}{\omega_3}}=e^{-\tilde{\beta}\omega_2},
\end{equation}

\noindent and

\begin{equation}
q=e^{-\beta|b_2|}\longrightarrow \tilde{r}=e^{-2\pi i \frac{\omega_1}{\omega_3}}=e^{-\tilde{\beta} i |b_2|},
\end{equation}

\noindent with $\tilde{\beta}=2\pi i/\omega_3=2\pi/|b_1|$. As we can see, the
transformation does two things: it switches the parameters $|b_1|$ and $\omega_2$,
interchanging geometric information about the $\mathrm{S^1}$ and the $\mathrm{S^3}$.
Furthermore, it turns the real parameter $q$ into a complex one, corresponding
to a Hopf surface with a non-direct product metric. The parameter $p$, however,
is mapped from real to a real variable.
Conversely, it is also possible to transform a complex parameter into a real
one by simply replacing $|b_2|$ with $i|b_2|$ in the original expression.

In terms of the superconformal index, the transformation can be written schematically as

\begin{equation}\label{tindex}
 \mathcal{I}(e^{-\beta|b_1|},e^{-\beta|b_2|},y)\xrightarrow{\mathrm{SL(3,\mathbb{Z}})}
 \mathcal{I}(e^{-\tilde{\beta}\omega_2},e^{-\tilde{\beta}i|b_2|},\tilde{y}),\end{equation}

\noindent where $\tilde{y}$ stands for an arbitrary $\mathrm{SL(3,\mathbb{Z})}$ transformed set of fugacities that may include both flavor and $R$-symmetry. We shall give explicit expressions for  $\tilde{y}$ for the examples
considered below.
 The index on the right hand side is simply the index of the same theory placed on a Hopf surface with parameters exchanged as described above.
Note, however, that whereas the original Hopf surface was defined using two contractions
$z_1\to p z_1,\, z_2 \to q z_2$ with $|p|,|q|<1$, for the modular transformed
surface the non-contractive regime $|q|=1$ becomes admissible. This will result
in the fact that modular transformed indices become well-defined meromorphic
functions even for $|q|=1$, i.e. effectively we cover a wider domain of
values of the moduli.

Since the parameters the transformed index depends on are the same but are now placed in different combinations, it is natural to ask what happens to the transformed index in a particular limit of the original one. For example,
one may consider the low ``temperature'' limit, i.e. $\beta\rightarrow\infty$, which corresponds to $\omega_2\rightarrow 0$, or $p\to 0$ and $q\to 0$. This limit was considered in \cite{SV2},
where it was indicated that for an appropriate choice of the flavour group fugacities
the indices reduce to the Hilbert series that counts gauge invariant operators \cite{HM}.
In the study of $\mathcal{N}=2$ indices in \cite{GRRY1}
it was called the Hall-Littlewood limit.

As mentioned above, this is the limit in which the Casimir energy is defined. In the transformed index, the only fugacity that depends on $\omega_2$ is $p=\mathrm{exp}(-\tilde\beta\omega_2)$, which behaves as $p\rightarrow 1$.
This limit is diverging and is not well understood on its own. However, the ``high temperature'' limit
for indices, $\beta\rightarrow 0$, or equivalently $\omega_2\rightarrow \infty$, or  $p \to 1$ and  $q\to 1$,
is well known. As noticed first in \cite{DSV}, in this case the indices reduce to partition
functions of three-dimensional field theories up to a diverging exponential, which was
shown in \cite{DPK} to be related to anomaly coefficients in the combination $a-c$.
The results of \cite{SV} were applied also in the context of formal holomorphic block
factorisations of $4d$ superconformal indices \cite{NP} with the corresponding
$\beta\rightarrow 0$ $3d$-reduction.
A more detailed consideration of this limit for several different theories was given in
\cite{ardehali}.
Analytically, the limit $p, q\rightarrow 1$ reduces
the elliptic hypergeometric integrals to the hyperbolic integrals.

We may also ask what happens for different $SL(3,\mathbb{Z})$ transformations, for example the combinations $\omega_3\rightarrow -\omega_1$ and $\omega_1\rightarrow \omega_3$ or $\omega_1\rightarrow -\omega_2$ and $\omega_2\rightarrow \omega_1$. In the former case, we get the transformations $p\rightarrow \mathrm{exp}(\beta|b_2|)$ and $q\rightarrow \mathrm{exp}(-\beta|b_1|)$, while in the latter, we get $p\rightarrow \mathrm{exp}(\beta' i|b_1|)$ and $p\rightarrow \mathrm{exp}(-\beta' \omega_2)$, with $\beta'=2\pi/|b_2|$. It is straightforward to see how the transformed parameters behave in the $\beta\rightarrow 0$ and $\beta\rightarrow \infty$ limits. However, none of these additional transformations will lead to the supersymmetric Casimir energy in the way outlined below.

\subsection{The recipe and a conjecture}

The $\mathrm{SL(3,\mathbb{Z})}$ transformation properties of the kernels of superconformal indices were
applied in \cite{SV} in the explanation of `t Hooft anomaly matching conditions for
Seiberg duality of $\mathcal{N}=1$ SQCD. It was based on the equivalence of the transformed indices
for both electric and magnetic theories. To this end, the transformed index has to be rewritten with the help of Eq. (\ref{gmod}) as the product of an exponential containing Bernoulli polynomials and a so-called modified
elliptic hypergeometric integral:

\begin{equation}
\mathcal{I}(\tilde{p},\tilde{r},\tilde{y})=e^{\varphi}I^{\mathrm{mod}}(\tilde{p},\tilde{r},\tilde{y}).
\end{equation}

\noindent The structure of the modified integral is essentially the
same as that of the superconformal index,
with elliptic gamma functions replaced by their modified counterparts
and the prefactor in front
of the integral slightly modified. To see this explicitly, consider Section 4 for a number of concrete examples. A crucial point is that the dependence of the exponent on the integration variables vanishes, i.e. it is independent of fugacities of the gauge group. While it is a priori not guaranteed, from the physical point of view it is a consequence of the absence of gauge
anomalies while from the mathematical point of view, it is the consequence of the balancing condition needed for the
original definition of elliptic hypergeometric integrals themselves \cite{spitheta}. This is true for all
dualities considered in \cite{SV2}. Matching of global anomalies is
in general equivalent to the condition that the exponent $\varphi$ matches for dual theories. In light of the recent discovery of the relationship between the anomaly polynomial and the Casimir energy through an equivariant integral, it is natural to conjecture that the Casimir energy is contained in $\varphi$. This is precisely what we find. The function $\varphi$ consists of a term proportional to $\tilde{\beta}$, which agrees with the Casimir energy, and a residual function $\cal R(\tilde{\beta})$ that contains terms subleading in $\tilde{\beta}$. This leads us to propose the following recipe for calculating the Casimir energy:

\begin{equation}\label{modrel}
\mathcal{I}(p,q,y)\xrightarrow{\mathrm{SL(3,\mathbb{Z}})}\mathcal{I}(\tilde{p},\tilde{r},\tilde{y})=e^{\tilde{\beta} E_{\mathrm{Casimir}}+\mathcal{R}(\tilde{\beta})}I^{\mathrm{mod}}(\tilde{p},\tilde{r},\tilde{y}).
\end{equation}

\noindent In words, there are three steps: i) transform the superconformal index
according to Eq. (\ref{transformation}) and Eq. (\ref{tindex}), ii) rewrite it
with the help of Eq. (\ref{gmod}) and iii) pull out the exponential factor and
identify the leading order term as the Casimir energy. Even though we do not
prove this recipe in full generality, we confirm it for several different
theories in the next section.

Inspecting Eq. (\ref{modrel}), one can see that upon bringing the leading order
exponential to the other side of the equation yields precisely the form of the
right hand side of Eq. (\ref{rel}). This leads us to conclude that the
partition function of the corresponding $\mathrm{SL(3,\mathbb{Z})}$ transformed
theory can be written as

\begin{equation}
\mathcal{\tilde{Z}}_{\mathrm{SUSY}}=e^{\mathcal{R}(\tilde{\beta})}
I^{\mathrm{mod}}(\tilde{p},\tilde{r},\tilde{y}).
\end{equation}

\noindent Since we could have also started the recipe with a transformed index, we get

\begin{equation}
\mathcal{Z}_{\mathrm{SUSY}}=e^{\mathcal{R}(\beta)}I^{\mathrm{mod}}.
\end{equation}

We conjecture that the recipe can be applied to all theories for which
the partition function can be written in the form of Eq. (\ref{rel}).
The partition function can then be
rewritten in terms of a modified elliptic hypergeometric integral introduced in \cite{DS}.

\section{Examples}

In this section, we give several examples for the application of the above recipe.

\subsection{$\mathbf{\mathcal{N}=1}$ SQCD}

Supersymmetric QCD with $\mathcal{N}=1$ superalgebra is a gauge theory with gauge group $\mathrm{SU(N_c)}$, flavor symmetry $\mathrm{SU(N_f)_L\times SU(N_f)_R\times U(1)_B}$ and R-symmetry $\mathrm{U(1)_R}$. We have summarized the matter content in Table \ref{matter}. In its conformal window, i.e. for $3N_c/2< N_f<3N_c$, it is subject to Seiberg duality \cite{Seiberg}. The superconformal index, which serves as a check for this duality, is given by

\begin{equation}\label{prefactor}
\mathcal{I}_{\mathrm{SQCD}}(p,q,y)=\kappa_{N_c}\int_{\mathbb{T}^{N_c-1}}
\frac{\prod_{j=1}^{N_c}
\prod_{l=1}^{N_f}\Gamma(s_lz_j,t_l^{-1}z_j^{-1};p,q)}{\prod_{1\leq j<k\leq N_c}\Gamma(z_jz_k^{-1},z_j^{-1}z_k;p,q)}\; \prod_{i=1}^{N_c-1}
\frac{dz_i}{2\pi i z_i},
\end{equation}

\noindent where we use the notation $\Gamma(a,b;p,q):=\Gamma(a;p,q)\Gamma(b;p,q)$ and $\kappa_{N_c}=\phi(p)^{N_c-1}\phi(q)^{N_c-1}/N_c!$. To see the transformation behavior, we rewrite the fugacities satisfying the balancing condition $\prod_{k=1}^{N_f}s_k t^{-1}_k=(pq)^{N_f-N_c}$ as $s_l=\mathrm{exp}(2\pi i \sigma_l /\omega_2)$, $t_l=\mathrm{exp}(2\pi i \tau_l /\omega_2)$ and $z_l=\mathrm{exp}(2\pi i u_l /\omega_2)$. Then, their transforms $\tilde{s}_l$, $\tilde{t}_l$ and $\tilde{z}_l$ are related simply by $\omega_2\rightarrow-\omega_3$.

Using Eq. (\ref{gmod}), the transformed index $\mathcal{I}_{\mathrm{SQCD}}(\tilde{y},\tilde{p},\tilde{r})$ can now be rewritten in the form indicated in Eq. (\ref{modrel}), with the Casimir energy $E_{\mathrm{Casimir}}$ given by

\begin{align}\label{sqcdenergy}
E_{\mathrm{Casimir}}=\lim_{\omega_3\to 0}\frac{\omega_3}{6}(\sum_{i=1}^{N_f}\sum_{j=1}^{N_c}(B_{3,3}(\sigma_i+u_j,\omega)+B_{3,3}(-\tau_i-u_j,\omega))\nonumber\\
-\sum_{1\leq i<j\leq N_c}(B_{3,3}(u_i-u_j,\omega)+B_{3,3}(u_j-u_i,\omega)))\nonumber\\
+\frac{N-1}{24}(\omega_1+\omega_2),
\end{align}

\begin{table}[t]
\centering
\begin{tabular}{lccccc}
\toprule
 &  $\mathrm{SU(N_c)}$  & $\mathrm{SU(N_f)}$ & $\mathrm{SU(N_f)}$ & $\mathrm{U(1)_B}$ & $\mathrm{U(1)_R}$\\
\toprule
$\mathbf{Q}^i$ & $f$ & $f$ & $1$ & $1$ & $(N_f-N_c)/N_f$ \\
$\mathbf{\tilde{Q}}_i$ & $\bar{f}$  &  $1$ & $\bar{f}$ & $-1$ & $(N_f-N_c)/N_f$ \\
$\mathbf{V}$ & $\mathrm{adj}$  & $1$ & $1$ & $0$ & $1$ \\
\toprule
\end{tabular}
\caption{The matter content of $\mathcal{N}=1$ SQCD. $f$ denotes the fundamental, $\bar{f}$ the antifundamental and $\mathrm{adj}$ the adjoint representation.}
\label{matter}
\end{table}

\noindent in complete agreement with the result of \cite{BBK}. Notice that the dependence of Eq.~\eqref{sqcdenergy} on the integration variables $u_i$ drops once we impose the balancing condition, which in the additive notation
should be chosen precisely as $\sum_{i=1}^{N_f}(\sigma_i-\tau_i)=(N_f-N_c)\sum_{k=1}^3\omega_k$
(the transition from the multiplicative to additive notation is slightly ambiguous due to the existence
of a natural period for the exponential function). Such a property serves as a criterion for the absence of
gauge anomalies \cite{SV}.

The residual function $\mathcal{R}(\tilde{\beta})$ reads

\begin{align}
\mathcal{R}(\tilde{\beta})=&-\frac{i \pi N_c}{3}\sum_{i=1}^{N_f}\left(C(\sigma_i)+C(-\tau_i)\right)
+\frac{i\pi(N_c^2-1)}{3}C(0)\,,
\end{align}

\noindent where we have defined\,\footnote{Notice that $C(x)$ is invariant under $x\rightarrow \alpha x$, $\omega_i\rightarrow \alpha \omega_i$, so it only depends on two ratios of three quasiperiods $\omega_i$.}

\begin{equation}
C(x)=B_{3,3}(x)-\frac{1}{\omega_3}\lim_{\omega_3\to0}\omega_3B_{3,3}(x).
\end{equation}

\noindent Finally, the modified elliptic hypergeometric integral is given by

\begin{equation}
\mathcal{I}^\mathrm{mod}_{\mathrm{SQCD}}=\kappa_{N_c}^{\mathrm{mod}}\int_{-\frac{\omega_3}{2}}^{\frac{\omega_3}{2}}\frac{\prod_{j=1}^{N_c}
\prod_{l=1}^{N_f}\mathcal{G}(\sigma_i+u_j,-\tau_i-u_j;\omega)}{\prod_{1\leq j<k\leq N_c}\mathcal{G}(u_i-u_j,u_j-u,i;\omega)}\prod_{k}^{N_c-1}\frac{du_k}{\omega_3},
\end{equation}

\noindent with

\begin{equation}\label{ka}
\kappa_{N_c}^{\mathrm{mod}}=\frac{(2\kappa_{2}^{\mathrm{mod}})^{N_c-1}}{N_c!},\quad
\kappa_{2}^{\mathrm{mod}}:= -\frac{\omega_3}{\omega_2}
\frac{\phi(p)\phi(q)\phi(r)}{2\phi(\tilde{q})}.
\end{equation}

\noindent  Evidently, this integral is well defined for $|q|=1$ \cite{DS}.
 The prefactor \eqref{ka} arises from Eq.~(\ref{prefactor}) in such a way
that the Casimir energy comes out correctly in the absence of flavor
symmetries. While this choice seems to be arbitrary at this point, the same
pattern also appears in all the other cases we have checked.

\subsection{$\mathcal{N}=1$ $\mathrm{SP(2N)}$ theory
with $\mathrm{SU(8)\times U(1)}$ flavour symmetry}

\begin{table}[t]
\centering
\begin{tabular}{lccccc}
\toprule
 &  $\mathrm{SP(2N)}$  & $\mathrm{SU(8)}$ & $\mathrm{U(1)}$ & $\mathrm{U(1)_R}$\\
\toprule
$\mathbf{Q}$ & $f$ & $f$ & $-\frac{N-1}{4}$  & $\frac{1}{4}$ \\
$\mathbf{X}$ & $T_A$  &  $1$ & $1$ & $0$ \\
$\mathbf{V}$ & $\mathrm{adj}$  & $1$ & $0$ & $\frac{1}{2}$ \\
\toprule
\end{tabular}
\caption{The matter content of the $\mathcal{N}=1$ theory with $\mathrm{SP(2N)}$ gauge symmetry and $\mathrm{SU(8)\times U(1)}$ flavour symmetry. $T_A$ denotes the antisymmetric tensor representation.}
\label{matter2}
\end{table}

In \cite{SV4}, Vartanov and one of the present authors studied a network
of dualities for a certain set of $\mathcal{N}=1$ theories. The starting point
is an ``electric" model described by SQCD-like theory with symplectic
$\mathrm{SP(2N)}$ gauge symmetry and
$\mathrm{SU(8)\times U(1)}$ flavour symmetry. Its field content is summarized
in Table \ref{matter2}. The corresponding superconformal index is given by

\begin{equation} \label{indSP2N}
\mathcal{I}^{SP(2N)}=\psi_N\int \prod_{1\leq j<k\leq N}
\frac{\Gamma(t z_j^{\pm 1} z_k^{\pm 1};p,q)}
{\Gamma(z_j^{\pm 1} z_k^{\pm 1};p,q)}\prod_{j=1}^N\frac{\prod_{k=1}^8
\Gamma(t^{\frac{1-N}{4}}( pq)^{\frac{1}{4}}y_k z_j^{\pm 1};p,q)}
{\Gamma(z_j^{\pm 2};p,q)}\frac{dz_j}{2\pi i z_j},
\end{equation}

\noindent where

\begin{equation}
\psi_N=\frac{\phi(p)^N\phi(q)^N}{2^N N!} \Gamma(t; p, q)^{N-1},
\end{equation}

\noindent with the fugacities $y_j\equiv\mathrm{exp}(2\pi i \alpha_j/\omega_2)$
and $t\equiv\mathrm{exp}(2\pi i \tau/\omega_2)$
satisfying the balancing condition

\begin{equation}
t^{2N-2}\prod_{j=1}^8 t_j=(pq)^2.
\end{equation}

\noindent Note that $\Gamma(a z^{\pm 1};p,q):=\Gamma(az,az^{-1};p,q)$.

The resulting expression for the Casimir energy is

\begin{eqnarray*} &&
E_{\mathrm{Casimir}}= \frac{N}{24}(\omega_1+\omega_2)+
\lim_{\omega_3\to 0}\frac{\omega_3}{6}(\sum_{1\leq j<k\leq N}(B_{3,3}(\tau\pm u_j\pm u_k)
-B_{3,3}(\pm u_j \pm u_k))
\\  &&  \makebox[2em]{}
+\sum_{j=1}^N(\sum_{k=1}^8 B_{3,3}(\frac{1-N}{4}\tau+\frac{\omega_1+\omega_3}{4}+
\alpha_k\pm u_j)-B_{3,3}(\pm 2u_j)) +(N-1)B_{3,3}(\tau)).
\end{eqnarray*}

\noindent Terms containing factors of $u_j$ again vanish completely due to the
balancing condition, corresponding to the absence of gauge anomalies.
The residual function is given by terms in higher powers of $\omega_3$.
The modified elliptic hypergeometric integral for this case can be read off
from the expression given in the end of the next section.

\subsection{$\mathcal{N}=1$ $\mathrm{E_7}$ flavor enhanced theories}

In \cite{DG}, Dimofte and Gaiotto constructed an $\mathcal{N}=1$ theory that possesses a point in its moduli space where flavour symmetry is enhanced to $\mathrm{E_7}$. This can be done by deforming a product of two copies of an $\mathrm{SU(2)}$ gauge theory $\mathcal{T}$ with $\mathrm{SU(8)}$ flavour symmetry. The superconformal index of one copy is given by

\begin{equation}
\mathcal{I}_{\mathcal{T}}(y)=\frac{\phi(p)\phi(q)}{2}
\int\frac{\prod_{i=1}^8 \Gamma((pq)^{1/4}y_i z^{\pm 1};p,q)}{\Gamma(z^{\pm 2};p,q)}\frac{dz}{2\pi i},
\end{equation}

\noindent where the fugacities $y_j\equiv\mathrm{exp}(2\pi i \alpha_j/\omega_2)$
satisfying $\prod_{i=1}^{8}y_i=1$ correspond to the $\mathrm{SU(8)}$
flavour symmetry.
This index possesses a nontrivial symmetry transformation described
by the the Weyl group $W(E_7)$  \cite{spitheta}. The model of \cite{DG}
with enhanced $\mathrm{E_7}$ flavor symmetry has now the index given by the product

\begin{equation}
\mathcal{I}_{\mathrm{E_7}}=\mathcal{I}_{\mathcal{T}}(y)\times \mathcal{I}_{\mathcal{T}}(y^{-1}).
\end{equation}

\noindent The Casimir energy can now be calculated in a straightforward manner, with the result

\begin{equation}
E_{\mathrm{Casimir}}=\frac{\omega_1+\omega_2}{\omega_{1}\omega_{2}}\left[\frac{1}{4}(\omega_1^2+\omega_2^2)+\frac{\omega_1\omega_2}{24}-\frac{1}{2}\sum_{i=1}^8\alpha_i^2\right],
\end{equation}

\noindent and the residual function

\begin{equation}
\mathcal{R(\tilde{\beta})}=\frac{\pi i}{3}\left[\frac{4\omega_3^2}{\omega_1\omega_2}+\frac{1}{4\omega_1\omega_2}\left(13\omega_3(\omega_1+\omega_2)+7\omega_1^2-9\omega_1\omega_2+7\omega_2^2-24\sum_{i=1}^8\alpha_i^2\right)\right].
\end{equation}

\noindent The corresponding modified elliptic hypergeometric integral is given by

\begin{equation}
\mathcal{I}^\mathrm{mod}_{\mathrm{E_7}}=\left(\kappa_2^{\mathrm{mod}}\right)^2
\int_{-\frac{\omega_3}{2}}^{\frac{\omega_3}{2}}
\frac{\prod_{j=1}^{8} \mathcal{G}(\frac{1}{4}(\omega_1+\omega_3)+\alpha_j \pm u,
\frac{1}{4}(\omega_1+\omega_3)-\alpha_j\pm u';\omega)}
{\mathcal{G}(\pm 2u,\pm 2u';\omega)}\frac{du}{\omega_3}\frac{du'}{\omega_3}.
\end{equation}

In \cite{DG} another model with the extended $E_7$ flavor symmetry was
suggested on the basis of a $5d$ chiral hypermultiplet interacting
with a $4d$ theory living on codimension 1 space. It is not clear how to compute
the Casimir energy for this system using our approach, since the
corresponding half-indices do not obey clear modular transformation
properties. However, using a similar idea, in \cite{LMP2014}
a $6d/4d$ theory was suggested where a $6d$ chiral hypermultiplet
was interacting with the $4d$ model described in the previous section which
lives in a ``corner" of the $6d$ space. The corresponding superconformal
index has the form

\begin{equation}
I_{6d/4d}=\frac{\mathcal{I}^{SP(2N)}}
{\prod_{1\leq j<k \leq 8}\Gamma(t^{\frac{N+1}{2}} (pq)^{\frac{1}{2}} y_j y_k ; p, q, t)},
\end{equation}

\noindent where $\mathcal{I}^{SP(2N)}$ is given in \eqref{indSP2N}.

The partition function of $6D$ chiral hypermultiplet has not yet been computed using
the localization method. However, we may speculate that the result is given by the modified superconformal
index proposed in \cite{LMP2014}, in analogy with the relation between $4d$ indices and
partition functions considered in \cite{ACM}-\cite{BBK} and the present work.
It is is given by

\begin{equation}
I_{6d/4d}^{\mathrm{mod}}=\psi_{6d/4d}^{\mathrm{mod}}
\int_{-\frac{\omega_3}{2}}^{\frac{\omega_3}{2}}
\prod_{1\leq i<j\leq N}\frac{\mathcal{G}(\tau \pm u_i\pm u_j;\omega)}
{\mathcal{G}(\pm u_i \pm u_j;\omega)}
\prod_{j=1}^N\frac{\prod_{k=1}^8\mathcal{G}(\alpha_k\pm u_j;\omega)}
{\mathcal{G}(\pm 2 u_j;\omega)}\frac{du_j}{\omega_3},
\end{equation}
where $\mathcal{G}(u;\omega):=\mathcal{G}(u;\omega_1,\ldots,\omega_3)$ and
the prefactor has the form for $\tau\equiv\omega_4$:

\begin{equation}
\psi_{6d/4d}^{\mathrm{mod}}=\frac{\left(\kappa_{2}^{\mathrm{mod}}\right)^N}{ N!}
\frac{ G(\omega_4;\omega_1,\ldots, \omega_3)^{N-1}  }
{ \prod_{1\leq j<k\leq 8}
\mathcal{G}(\frac{N}{4}\omega_4 +\frac{1}{4}\sum_{i=1}^4\omega_i+\alpha_j+\alpha_k;
\omega_1,\ldots,\omega_4)}
\end{equation}
with a product of modified elliptic gamma functions in the denominator.
Applying the logical line of the previous considerations, we can expect
that the supersymmetric Casimir energy of the described $6d/4d$ system
will be given by

\begin{equation}
E_{\mathrm{Casimir}}^{6d/4d}=E_{\mathrm{Casimir}}^{\mathrm{SP(2N)}} -\lim_{\omega_3\to 0}\frac{\omega_3}{24}
\sum_{1\leq j<k\leq 8}B_{4,4}(\frac{N}{4}\omega_4
+\frac{1}{4}\sum_{i=1}^4\omega_i+\alpha_j+\alpha_k),
\end{equation}

\noindent which receives now also contributions
from the Bernoulli polynomial of fourth order, as it is given in section 2.1.

\subsection{$\mathcal{N}=2$ strongly coupled $\mathrm{E_6}$ SCFT}

Argyres-Seiberg duality \cite{AS} maps an $\mathcal{N}=2$ supersymmetric
$\mathrm{SU(3)}$ gauge theory with $N_f=6$ flavour symmetry with a weakly
coupled description to a theory that involves a strongly coupled sector
without a Lagrangian description. This sector is a superconformal theory
with an $\mathrm{E_6}$ flavour symmetry \cite{MN}. While the superconformal index
of such a theory cannot be written down by conventional means, the
problem was circumvented in \cite{GRRY2} in an elegant manner by employing
 mathematical techniques developed in \cite{SW}.
The resulting index can be written as

\begin{align}
\mathcal{I}_{\mathrm{E_6}}=&\frac{1}{2}\frac{\phi(p)\phi(q)}{\Gamma\left(\hat{r}(\hat{s}^4\hat{t}^{-2})^{\pm 1}\right)}\int_\mathbb{T} \frac{ds}{s}\frac{\Gamma\left((\hat{s}\hat{t})^{-\frac{1}{2}}(\hat{s}^2\hat{t}^{-1})^{\pm 1} s^{\pm 1}\right)}{\Gamma\left((\hat{s}\hat{t})^{-1},s^{\pm 2}\right)}\mathcal{I}(s,r,\mathbf{y},\mathbf{z})\nonumber\\
+&\frac{1}{2}\frac{\Gamma\left(\hat{t}^2 \hat{s}^{-4}\right)}{\Gamma\left(\hat{r}\hat{t}^2 \hat{s}^{-4}\right)}\left[\mathcal{I}\left(s=\left(\hat{s}\hat{t}^{-1}\right)^{\frac{3}{2}},r,\mathbf{y},\mathbf{z}\right)+\mathcal{I}\left(s=\left(\hat{s}^{-1}\hat{t}\right)^{\frac{3}{2}},r,\mathbf{y},\mathbf{z}\right)\right]\nonumber\\
+&\frac{1}{2}\frac{\Gamma\left(\hat{t}^{-2} \hat{s}^{4}\right)}{\Gamma\left(\hat{r}\hat{t}^{-2} \hat{s}^{4}\right)}\left[\mathcal{I}\left(s=\hat{s}^{-\frac{5}{2}}\hat{t}^{\frac{1}{2}},r,\mathbf{y},\mathbf{z}\right)+\mathcal{I}\left(s=\hat{s}^{\frac{5}{2}}\hat{t}^{-\frac{1}{2}},r,\mathbf{y},\mathbf{z}\right)\right],
\end{align}

\noindent where $\Gamma(z):=\Gamma(z;p,q)$ and
$\mathcal{I}(s,r,\mathbf{y},\mathbf{z})$ is the index of a weakly coupled $\mathrm{SU(3)}$ theory with six hypermultiplets, and is given by

\begin{align}
\mathcal{I}(s,r,\mathbf{y},\mathbf{z})=\frac{\phi(p)^2\phi(q)^2}{6}\Gamma(\hat{r})^2&\int \prod_{i,j=1}^3\Gamma\left((\hat{s}\hat{t})^{\frac{1}{2}}\left(s^{\frac{1}{3}}\frac{\hat{z}_i}{x_j}\right)^{\pm 1},(\hat{s}\hat{t})^{\frac{1}{2}}\left(s^{-\frac{1}{3}}\hat{y}_ix_j\right)^{\pm 1}\right)\nonumber\\
&\times \prod_{i\neq j}\frac{\Gamma(\hat{r}x_i x_j^{-1})}{\Gamma(x_i x_j^{-1})}
\prod_{i=1}^{2}\frac{dx_i}{2\pi i x_i}.
\end{align}

\noindent The fugacities $\hat{r}\equiv\mathrm{exp}(2\pi i \rho/\omega_2)$, $\hat{s}\equiv\mathrm{exp}(2\pi i \sigma/\omega_2)$ and $\hat{t}\equiv\mathrm{exp}(2\pi i \tau/\omega_2)$ satisfy the balancing condition $\hat{r}\hat{s}\hat{t}=pq$ and are related to the parameters $t$, $v$ and $w$ in \cite{GRRY2} by $\hat{r}=t^2v$, $\hat{s}=t^{\frac{8}{3}}v^{-\frac{2}{3}}w^{-\frac{1}{3}}$. Furthermore, we have $\hat{y}_j\equiv \mathrm{exp}(2\pi i \alpha_j/\omega_2)=\tilde{r}^{-1}y_j$ and $\hat{z}_j\equiv\mathrm{exp}(2\pi i \beta_j/\omega_2)=\tilde{r}^{-1}z_j$, where $\prod_{j=1}^3x_j=\prod_{j=1}^3y_j=\prod_{j=1}^3z_j=1$ and $\tilde{r}\equiv\mathrm{exp}(2\pi i r/\omega_2)$.

Applying our $\mathrm{SL(3,\mathbb{Z})}$ method as in the examples above yields an expression for the Casimir energy,

\begin{align}
E_{\mathrm{Casimir}}=\frac{-\rho}{\omega_1\omega_2}\Bigg[&\frac{3}{2}\sum_{i=1}^3\left(\alpha_i^2+\beta_i^2\right)+\frac{71}{6}\rho^2-\frac{91}{4}(\omega_1\rho+\omega_2\rho-\omega_1\omega_2)\nonumber\\
&+\frac{131}{12}(\omega_1^2+\omega_2^2)-36(\omega_1\tau+\omega_2\tau-\rho\tau)+27\tau^2\Bigg],
\end{align}

\noindent which is in precise agreement with the results of \cite{BBK}. This can be seen after the identifications $\rho=-\sigma$, $\tau=\frac{1}{3}(-e+2(\omega_1+\omega_2+\omega_3 +\sigma)$, $\alpha_j=y_j-r$ and $\beta_j=z_j-r$, where we have used the notation of \cite{BBK}. It is again straightforward to write down the residual function and the modified elliptic hypergeometric integral, but we refrain from doing so explicitly for the sake of brevity.

\subsection{$\mathbf{\mathcal{N}=4}$ supersymmetric Yang-Mills theory}

 As a final example, we consider $\mathcal{N}=4$ supersymmetric Yang-Mills theory with $\mathrm{SP(2N)}$ gauge group. One has to start with a more general expression than \eqref{index0}, see \cite{KMMR}. The explicit expression \cite{GPRR,SV3} is given by

\begin{equation}
\mathcal{I}_{\mathrm{SP(2N)}}=\chi_N\int \prod_{1\leq i<j\leq N}\frac{\prod_{k=1}^3\Gamma(s_k z_i^{\pm 1}z_j^{\pm 1};p,q)}{\Gamma(z_i^{\pm 1}z_j^{\pm 1};p,q)}
 \prod_{i=1}^N\frac{\prod_{k=1}^3\Gamma(s_k z_i^{\pm 2};p,q)}{\Gamma(z_i^{\pm 2})}\frac{dz_i}{2\pi i z_i},
\end{equation}

\noindent with $s_k=\mathrm{exp}(2\pi i \alpha_k/\omega_2)$ and

\begin{equation}
\chi_N=\frac{\phi(p)^N\phi(q)^N}{2^N N!}\prod_{k=1}^3\Gamma(s_k;p,q)^N.
\end{equation}

\noindent The balancing condition reads $s_1s_2s_3=pq$. Applying the same transformation as above leads, in complete analogy, to

\begin{equation}
E_{\mathrm{Casimir}}=(2N^2+N)\frac{\alpha_1\alpha_2\alpha_3}{2\omega_1\omega_2},
\end{equation}

\noindent while the residual function takes the simple form

\begin{equation}
\mathcal{R}(\tilde \beta)=\frac{3\pi i N}{2}.
\end{equation}

\noindent The modified elliptic hypergeometric integral is given by

\begin{equation}
I_{\mathcal{N}=4}^{\mathrm{mod}}=\chi_N^{\mathrm{mod}}\int_{-\frac{\omega_3}{2}}^{\frac{\omega_3}{2}} \prod_{1\leq i<j\leq N}\frac{\prod_{k=1}^3\mathcal{G}(\alpha_k\pm u_i\pm u_j;\omega)}{\mathcal{G}(\pm u_i \pm u_j;\omega)}
\prod_{i=1}^N\frac{\prod_{k=1}^3\mathcal{G}(\alpha_k\pm 2u_i;\omega)}{\mathcal{G}(\pm 2 u_i;\omega)}\frac{du_i}{\omega_3},
\end{equation}

\noindent with

\begin{equation}
\chi_N^{\mathrm{mod}}=\frac{1}{2^N N!}\left(-\frac{\omega_3}{\omega_2}\right)^{N}\left(\frac{\phi(p)\phi(q)\phi(r)}{\phi(\tilde{q})}\right)^{N}\prod_{k=1}^3\mathcal{G}(\alpha_k;\omega)^N.
\end{equation}

\noindent and $\sum_{k=1}^3\alpha_k=\sum_{i=1}^3\omega_i$. We have also performed the calculation for all the other classical groups, with the result being

\begin{equation}
E_{\mathrm{Casimir}}=d_G\frac{\alpha_1\alpha_2\alpha_3}{2\omega_1\omega_2},
\end{equation}

\noindent where $d_G$ is the dimension of the group. This is again in full agreement
with \cite{BBK}. The residual function is the same in all cases.

\section{Discussion}

In this article, we have studied the connection between $\mathrm{SL(3,\mathbb{Z})}$
transformations,  the supersymmetric partition function and the superconformal
index of four-dimensional theories.
We have proposed a new recipe to extract the supersymmetric Casimir energy
from the index alone, and confirmed it for several theories with $\mathcal{N}=1,2,4$
supersymmetry. We have also predicted the Casimir energy for two very
interesting $\mathcal{N}=1$ theories. Moreover, we have shown that the
structure of Eq. (\ref{rel}) emerges from $\mathrm{SL(3,\mathbb{Z})}$
transformations and, given that the localization result holds, found a way
to write the partition function in terms of a modified elliptic hypergeometric integral.
It is tempting to state that the modified elliptic hypergeometric
integrals actually coincide with partition functions, as the exponent in the computations of the latter
for example in the case of a chiral superfield in \cite{CS} is similar to the $\mathrm{SL(3,\mathbb{Z})}$
transformation factor. However, due to the complicated
nature of the regularization procedure such a statement would require rigorous
mathematical justification (see \cite{ACM,ACDKLM,CS,ALS} for detailed considerations of this problem).

Finally, we want to comment on the geometric and physical interpretation of the  $\mathrm{SL(3,\mathbb{Z})}$
transformation and the emergence of the Casimir energy. Consider the action of the transformations $\omega_2\rightarrow-\omega_3$ and $\omega_3\rightarrow\omega_2$ on the identification of Eq. (\ref{hopf}). We see that  the resulting identification still gives a primary Hopf surface, but with different defining parameters. As the transformation is not continuously connected to the identity, it is a large diffeomorphism of the manifold. The failure of the superconformal index/partition function to be invariant under this diffeomorphism points towards the presence of a gravitational anomaly (see \cite{CS} for a similar phenomenon).

There are many open questions and avenues to pursue in the future,
some of which we want to mention in the following:
\begin{itemize}

\item It would be interesting to find a physical interpretation of the residual phase polynomial
$\mathcal{R}(\tilde{\beta})$. One can say that the original cocycle phase function emerging from the $\mathrm{SL(3,\mathbb{Z})}$ transformation
represents some kind of a modified Casimir energy whose low temperature
$\beta\to\infty$ leading term yields the true Casimir energy. According to
the considerations of \cite{NP}, one can interpret the expression \eqref{GMOD1}
as a combination of two superconformal indices defined on two different manifolds arising from a Heegard-decomposition of the original manifold.
Then it remains to clarify the meaning of the result of such a gluing
given by the expression \eqref{gmod}.

\item In section 3.2, we have shown that a particular $\mathrm{SL(3,\mathbb{Z})}$ transformation takes the $\beta\rightarrow\infty$ limit of the original index into the $p\rightarrow 1$ limit of the
    transformed index. It would be interesting to investigate analytically such a diverging limit.  The high temperature limit $\beta\rightarrow\infty$, with $p, q\to1$ taken simultaneously, the degeneration leads to three-dimensional theories. It remains to investigate in detail all regimes that can be reached by the  $\mathrm{SL(3,\mathbb{Z})}$ transformations.

\item In \cite{Razamat}, the $\mathcal{N}=2$ Schur index was modified in such a way that it is invariant under modular transformations. It is in principle conceivable that a similar modification exists for the superconformal index and $\mathrm{SL(3,\mathbb{Z})}$. We have not discussed such a modification in the present article.

\item There are other ways of computing the supersymmetric Casimir energy, e.g. as an equivariant integral over the anomaly polynomial \cite{BBK} or as a limit of an index-character counting twisted holomorphic modes \cite{MS}. Even though the relationship between $\mathrm{SL(3,\mathbb{Z})}$ and  anomaly coefficients was found already in \cite{SV}, it would be important to clarify what is the precise relation to these approaches.

\item In \cite{LV}, an intriguing connection between partition functions, topological strings and $\mathrm{SL(3,\mathbb{Z})}$ transformations was uncovered. It would be of interest to know if these insights are in any way related to our results.

\item Finally, it would be interesting to have, even in the absence of a proof, more explicit checks of the recipe. For example, it would be desirable to see that it works for other theories with $\mathcal{N}=2$ supersymmetry \cite{GPRR} or with more complicated superconformal indices, like the linear quivers of \cite{BS}.

\end{itemize}

\bigskip
The authors would like to thank H.-C.~Kim and J. Sparks for helpful discussions. V.S. is indebted to A. Kapustin for a discussion of
the structure of partition functions computed via localization. D.R. thanks the University of Wisconsin-Madison, and F.B. the Mathematical Institute of the University of Oxford for hospitality during the completion of this work. F.B. was supported by the Austrian Science Fund FWF, project no.  P26366, and the FWF doctoral program Particles \& Interactions, project no. W1252. D.R. was supported by a grant of the Max Planck Society. Results of Section 3 have been worked out within the Russian Science Foundation project no. 14-11-00598.
V.S. is partially supported by Laboratory of Mirror Symmetry NRU HSE, RF government grant,
ag. no. 14.641.31.0001.

\end{document}